# Part 1: Disruption of Water-Carbon Cycle under Wet Climate Extremes

**Maheshwari Neelam[1,2] and Christopher Hain[2]**

[1] Universities Space Research Association, Huntsville, AL 35801, USA

[2] NASA Marshall Space Flight Center, Earth Science Branch, Huntsville, AL 35801, USA

Corresponding author: Maheshwari Neelam (maheshwari.neelam@nasa.gov)

## Abstract

Modern climate change presents unprecedented challenges, posing critical crises that threaten sustainable development, human well-being, and planetary health. A significant concern is the potential for global warming to cause irreversible disruptions to the water-carbon cycle, a topic that remains underexplored. This study seeks to address a crucial knowledge gap by examining how increasing wet extremes impact ecosystem productivity. The research agenda focuses on three primary questions: 1) How do the intensity and duration of various wet extremes affect evapotranspiration across different watersheds and terrestrial biomes? 2) How do immediate and lagged responses to wet extremes vary across different biomes, and what insights do these temporal patterns provide about the causal and predictive relationships between wet extreme and evapotranspiration? 3) To what extent do watershed characteristics (such as soil properties, hydrological conditions, and vegetation factors) modulate the relationship between wet extremes and ecosystem productivity? As climate change alters precipitation patterns, understanding these complex ecosystem responses becomes crucial for developing adaptive strategies and improving food and water resource management.

**Significance statement:** This study addresses a critical knowledge gap in understanding the impact of increasing wet extremes on ecosystem productivity amid rapidly changing climate. By examining complex relationships between wet climate extremes and evapotranspiration across diverse watersheds and biomes, the research reveals multi-timescale interactions. It highlights both immediate and lagged effects of extreme precipitation on ecosystem productivity, identifying key watershed characteristics that modulate this coupling. The study quantifies thresholds in these characteristics and lagged effects, essential for developing climate-resilient management strategies. By integrating these insights into watershed management approaches, stakeholders can better adapt to and mitigate the impacts of escalating climate extremes on water resources and ecosystem productivity.

# 1. Introduction

Anthropogenic climate change is intensifying Earth's hydrologic cycle, leading to more frequent and severe extreme precipitation events (1) (2) (3) (4) (5). As global temperatures rise, the atmosphere's water-holding capacity increases, following the Clausius-Clapeyron relation, with approximately 7% more water vapor per 1°C of warming (6) (7) (8). Climate models project a robust increase in extreme precipitation globally and regionally, with heavy precipitation becoming more frequent and intense as warming progresses (9) (10) (11) (12). The consequences of these changes are far-reaching and complex. Extreme precipitation can have devastating direct societal impacts, including flooding (10) (13), soil erosion, and agricultural damage(14), as well as lagged impacts like pests/pathogen outbreaks (15), soil carbon loss (16), and changes in plant phenology (17). Altered rainfall distributions cause ecosystems with high rainfall to experience more frequent soil waterlogging events beyond traditional flood zones, particularly in lowland humid regions. Excessive rainfall can result in tree mortality due to soil waterlogging (18), with maladapted species experiencing higher mortality rates in saturated soils (19) (20). In the Arctic, increased summer precipitation is expected to continue, with a greater proportion falling as liquid rain rather than snow, potentially accelerating permafrost thaw and degradation (21), which could further contribute to significant greenhouse gas emissions (22). As such, many regions are witnessing shifts in their local water and carbon balance, placing substantial stress on both ecological and societal water and food resources. The IPCC SREX (23) special report defines climate extremes as occurrences where weather or climate variables exceed typical threshold values, either above the 90th or below the 10th percentile of their probability density function. The World Meteorological Organization (WMO) Commission for Climatology (CCl)/CLIVAR/JCOMM Expert Team on Climate Change Detection and Indices (ETCCDI) has developed a suite of indices to quantify climate extremes, providing a standardized framework for assessing changes in weather patterns. The average annual maximum precipitation amount in a day (RX1Day) has significantly increased since the mid-20th century over land (24) and in the humid and dry regions of the globe (25). The probability of precipitation exceeding 50 mm/day increased during 1961–2018 (26) (27). The globally averaged annual fraction of precipitation from days in the top 5% (R95P) has also significantly increased (24). Extreme climate events are characterized by their great magnitude and short duration compared to average climate conditions. Despite their brevity, these events have a disproportionately large impact on the environment (28). They act as powerful drivers of change in two ways: through their immediate occurrence and through the ecosystem's response to them. When assessing potential ecosystem impacts, these extreme events likely play a more crucial role than gradual climate trends (29). This is because the environmental response to a single extreme event can often be significant enough to alter ecosystem functioning (30). Consequently, extreme events may accelerate the effects of climate change beyond what we might expect from long-term trends alone (28).

These unprecedented changes in precipitation extremes have profound implications for ecosystem functioning, particularly in terms of productivity (31)(32)(33). However, the relationship between precipitation and ecosystem productivity is not straightforward, complicated by lag effects and nonlinear responses. This nonlinearity can manifest as concave-down curves, where productivity increases rapidly with initial precipitation but plateaus at higher levels, or concave-up curves, where productivity responds slowly at first but accelerates with increasing precipitation. Legacy effects further complicate this relationship, as dry or wet conditions in one year can influence productivity in subsequent years. For instance, the influence of rainfall on vegetation often accumulates over time, with the full benefits of a significant rainfall event potentially not realized for days or weeks. Recent evidence suggests that extreme precipitation events can trigger lagged responses of ecosystems, substantially affecting ecosystem productivity on a longer timescale (34)(35)(36)(37)(38). Evapotranspiration effectively links water and carbon cycles, serving as a simple way to capture ecosystem productivity (39) (40). Nonlinear changes in evapotranspiration are typically driven by variations in climatic, ecological, geological, and human activities, which reshape the distribution of surface energy, carbon and water within a watershed. Thus, evapotranspiration serves as a versatile tool, effectively representing ecosystem productivity through the interaction between the water, energy, and carbon nexus. As climate change continues to alter precipitation patterns, understanding these complex ecosystem responses becomes increasingly critical for developing adaptive strategies and improving our ability to forecast and manage food and water resources.

Building on this understanding of rapidly changing precipitation patterns and evapotranspiration's role in ecosystem dynamics, as well as the urgent need for adaptive strategies, our study aims to delve deeper into the specific impacts of precipitation extremes by considering follow hypothesis: (i) ecosystem productivity varies with the type and intensity of extremes, (ii) immediate and lagged responses to wet extremes varies by biome type, (iii) watershed characteristics, including soil properties, hydrological conditions, and vegetation factors, will significantly influence how wet extremes impact ecosystem productivity, affecting positive or negative outcomes. To test our hypotheses, we employ machine learning (ML) methods, specifically random forest (RF) and Shapley Additive Explanations (SHAP), for attribution of watershed characteristics. We also use cluster analysis to determine mean lagged behaviors of various precipitation extremes. Through a large-scale and long-term quantitative analysis, we derive specific thresholds for watershed characteristics to guide adaptation and remediation efforts in the face of climate extremes (41).

## 2. Data and Method

The analysis is conducted utilizing two different evapotranspiration products: Atmosphere-Land Exchange Inverse, ALEXI (2000-2023) (42) and Global Land Evaporation Amsterdam Model, GLEAM (1980-2023) (43), representing distinct methodological approaches. ALEXI is an energy balance model that uses thermal infrared data to estimate land surface temperature, while GLEAM is based on the Priestley-Taylor framework and incorporates multiple satellite data sources. Both products are analyzed at a monthly scale at a common spatial resolution of ~0.5 degrees. While ALEXI offers higher temporal resolution and potential for near-real-time estimates, GLEAM provides a longer historical record. The comparison of these two approaches allows for a comprehensive assessment of evapotranspiration dynamics across different ecosystems and climatic conditions. In this analysis, we use fractional evapotranspiration (fET) which is defined as the ratio of actual evapotranspiration to potential evapotranspiration. This dimensionless index allows for comparison across diverse climatic regions and vegetation types, effectively capturing the relative water stress and productivity of ecosystems. By accounting for both atmospheric demand and water availability, fET serves as a sensitive indicator of ecosystem response to climate variability and extremes.

### 2.1. Atmosphere-Land Exchange Inverse (ALEXI)

ALEXI is an energy-budget based algorithm that calculates evapotranspiration as the residual of net radiation by subtracting soil heat flux and sensible heat flux. It uses thermal infrared data to estimate surface temperature and couples a two-source energy balance (TSEB) model with an atmospheric boundary layer model. ALEXI is available at 0.05° x 0.05° and uses land surface temperature observations at two times during the morning to compute the energy balance. The model is less sensitive to biases in instantaneous satellite-based temperature estimates due to its time-differential approach. ALEXI can provide daily evapotranspiration estimates and has been implemented globally, with the capability to be downscaled to higher resolutions through the DisALEXI framework. However, since evapotranspiration estimates are based on thermal information, values used to calculate evapotranspiration depend on clear-sky conditions, meaning only a portion of the ALEXI modeling domain can be updated on any given day. To address this limitation, temporal compositing of clear-sky evapotranspiration is necessary to achieve full domain coverage. Compositing also helps reduce noise in evapotranspiration retrievals, primarily due to incomplete cloud clearing in the land surface temperature inputs to ALEXI. Typically, 28-day composites are generated using a 4-week moving window, with timestamps corresponding to the end date of each period. Anderson et al. (2011) (44) showed that the ALEXI-derived Evaporative Stress Index (ESI) effectively captured drought conditions, often providing earlier warning than precipitation-based indices. Cammalleri et al. (2014) (45) validated ALEXI evapotranspiration across European flux towers, finding good agreement with ground-based measurements (average correlations of 0.71 and relative errors around 28%).

## 2.2. Global Land Evaporation Amsterdam Model (GLEAM)

GLEAM uses a unique algorithmic approach based on the Priestley-Taylor framework to estimate global evapotranspiration. It incorporates multiple satellite data sources, including surface soil moisture, vegetation optical depth, and land surface temperature observations, along with ERA5 meteorological reanalysis data. GLEAM provides daily evapotranspiration estimates at a global scale with a spatial resolution of 0.25° x 0.25°. Unlike energy-budget based models, GLEAM does not calculate evapotranspiration as a residual of the energy balance. Instead, it uses a more direct estimation approach that integrates various satellite-derived inputs. GLEAM estimates land evaporation mainly based on parameterized physical processes, with stress conditions parameterized as a function of dynamic vegetation information and available water in the root zone. A key feature of GLEAM is its detailed parameterization of forest interception. GLEAM calculates different components of land evaporation separately, including transpiration, bare-soil evaporation, interception loss, open-water evaporation, and sublimation. This approach allows for a comprehensive representation of evapotranspiration processes across diverse ecosystems. Miralles et al. (2011) (46) found that GLEAM estimates correlated well with eddy-covariance measurements across various biomes ($R^2 = 0.83$). Martens et al. (2017) (47) reported that GLEAM v3 showed improved performance over its predecessors, with a mean correlation of 0.78 against in situ observations. However, GLEAM has some limitations. It tends to underestimate evapotranspiration in water-stressed conditions and overestimate in energy-limited environments. Despite these challenges, GLEAM offers comprehensive global long-term coverage.

## 2.3. Climate Extremes Indices (CEIs)

Over the past two decades, significant efforts have been made to assess changes in the spatial and temporal patterns of extreme climate events globally using observational climatic data (48) (49). One such dataset is NASA's Modern Era Retrospective Analysis for Research and Applications, version 2 (MERRA-2) (50) (51), which includes Monthly Extremes Detection

Indices at spatial resolution of 0.625° longitude by 0.5° latitude from 1980-2023 (52). Extreme detection indices in MERRA-2 are derived using daily precipitation or daily mean, maximum, or minimum 2-meter temperature. These indices provide long-term characteristics of extreme events, such as their intensity, duration, and frequency, using daily exceedances of percentile thresholds. The dataset consists of 15 wet extreme indices calculated based on daily exceedances of percentiles and aggregated into monthly indices. We use 14 indices related to precipitation extremes, Table 1.

**Table 1: Precipitation-Related CEIs:** The indices are calculated based on daily exceedances of the percentiles and aggregated into monthly indices representing extreme precipitation events.

| Acronym | Description | Units |
|---|---|---|
| **CWD** | **Consecutive Wet Days** - The maximum number of consecutive days with more than 1 mm of precipitation. | days |
| **R10MM** | **Heavy Rain Days** - The count of days with at least 10 mm of precipitation. | days |
| **R20MM** | **Very Heavy Rain Days** - The count of days with at least 20 mm of precipitation. | days |
| **R90D** | **Wet Days** - The count of days with precipitation above the 90th percentile. | days |
| **R90P** | **Wet Precipitation** - The total precipitation from days with rainfall above the 90th percentile. | mm day$^{-1}$ |
| **R95D** | **Very Wet Days** - The count of days with precipitation above the 95th percentile. | days |
| **R95P** | **Very Wet Precipitation** - The total precipitation from days with rainfall above the 95th percentile. | mm day$^{-1}$ |
| **R99D** | **Extremely Wet Days** - Count of days that exceed the 99th percentile of precipitation. | days |
| **R99P** | **Extremely Wet Precipitation** - Mean precipitation on days that exceed the 99th percentile of precipitation. | mm day$^{-1}$ |
| **RX1Day** | **Maximum One-Day Precipitation** - The highest precipitation amount recorded in a single day. | mm day$^{-1}$ |
| **RX5Day** | **Maximum Five-Day Precipitation** - The highest precipitation amount recorded over any five consecutive days. | mm day$^{-1}$ |
| **RX5Daycount** | **Five-Day Heavy Precipitation Periods** - The count of five-day periods with heavy precipitation. | count |
| **SDII** | **Simple Daily Intensity Index** - The ratio of total precipitation to the number of wet days within a year. | mm day$^{-1}$ |
| **WD** | **Wet Days** - The count of days with more than 1 mm of precipitation. | days |

### 2.4. Watershed Characteristics

HydroATLAS (53) provides a comprehensive, consistently organized, and fully global data compendium that presents a wide range of hydro-environmentally relevant characteristics at both sub-basin and river reach scale. This high-resolution dataset is derived from global digital elevation models (DEMs) and offers sub-basin characteristics at 12 different spatial scales, ranging from coarse (level 1) to detailed (level 12). At its highest level of subdivision (level 12), BasinATLAS contains 1.0 million sub-basins with an average area of 130.6 km$^2$. For this analysis, we consider the level 4 subdivision, which provides an appropriate scale for watersheds that aligns with MERRA-2 data resolution. The hydro-environmental attributes considered in the study are: River Area (RA), Groundwater Table Depth (GWT), Elevation (minimum, maximum, and average), average Slope, Landcover (Forest, Cropland, Pasture, Irrigated, Urban, Permafrost) Extent, Normalized Difference Vegetation Index (NDVI), Gross Primary Productivity (GPP), Clay %, Silt %, Sand %, and Soil Organic Carbon (SOC). These variables were selected based on two primary criteria: 1) the continuity and reliability of data across diverse watersheds, and 2) their significance in characterizing watershed dynamics and ecosystem responses to climate extremes. By incorporating these diverse yet interconnected variables, the study aims to provide a holistic understanding of how watersheds respond to and interact with precipitation extremes, thereby enhancing our ability to predict and manage ecosystem responses in the face of changing climate patterns .

### 2.5. Methodology

This analysis explores the relationships between Climate Extreme Indices (CEIs) and fractional evapotranspiration (fET) using Spearman and Kendall correlation coefficients ($p < 0.1$). To ensure robust results, only extreme events affecting at least 40% of the watershed are considered, and correlation values must exceed an absolute value of 0.3 with $p < 0.1$. The CEIs data were normalized using the Box-Cox transformation, although this did not significantly alter the final results. The purpose of applying these lags is to explore potential temporal relationships between climate extremes and evapotranspiration. By shifting the data series relative to each other, the analysis can identify whether there are delayed effects or precursor signals in the relationship between climate extremes and evapotranspiration. This approach allows for a comprehensive examination of how fET responds to or potentially influences climate extremes across different time scales, which is crucial for understanding the complex interactions between climate and hydrological processes. Precipitation-related CEIs exhibit complex and varied lagged effects on ecosystem processes. Thus, lagged correlations were computed for durations of ±6 months. Positive lags might represent delayed ecosystem responses to precipitation events (e.g., increased fET following rainfall). Conversely, negative lags could indicate anticipatory changes in fET before precipitation events or seasonal transitions. A lag of zero represents the simultaneous relationship between fET and CEIs. A cluster analysis using DBSCAN (Density-Based Spatial Clustering of Applications with Noise) algorithm (54) is further employed to determine mean lag times for causal and predictive relationships between CEIs and fET, identifying groups of similar correlation-lag combinations. This approach highlights whether certain CEIs consistently lead or lag fET changes and quantifies average time differences. The clustering process ensures sufficient data points (watersheds) >50 were used. The algorithm employs a grid search approach to find optimal DBSCAN parameters: epsilon and minimum samples. It iterates through a range of values for both parameters, creating DBSCAN models for each combination. The quality of each clustering is evaluated using the silhouette score, which measures how similar an object is to its own cluster compared to other clusters. This adaptive approach ensures the clustering is optimized for the specific characteristics of the data.

The coupling strength between CEIs and fET, determined through Spearman correlation, is further analyzed using a Random Forest (RF) model (55) and Shapley Additive Explanations (SHAP) (56). This approach quantifies the contribution of watershed characteristics in driving the coupling strength. RF, an ensemble learning method, was chosen for its versatility, power in regression tasks, and ability to mitigate overfitting. While XGBoost was also evaluated, RF demonstrated superior performance and efficiency. Both RF and Spearman correlation methods were selected for their robustness to outliers and skewed distributions. The RF model was optimized through hyperparameter tuning, focusing on parameters such as the number of estimators, maximum depth, minimum samples required to split, and learning rate. This optimization aimed to minimize the Root Mean Square Error (RMSE) while maximizing $R^2$ and overall accuracy. To address the interpretability limitations of RF, we employed SHAP, a game-theoretic approach that provides a consistent and interpretable metric for understanding the relative contribution of input features and their interactions. SHAP values were used to rank feature importance and elucidate the nature of these influences—whether positive or negative—on the model output. This method overcomes the limitations of traditional feature importance measures, offering a more nuanced and personalized interpretation. By averaging SHAP values at points where they change sign, inflection points were identified, quantifying thresholds for each significant watershed characteristic that influence the coupling strength between CEIs and fET.

## 3. Results and Discussion

Among the wet CEIs, we observe a higher frequency of intensity-based wet CEIs (e.g., R90P, R95P, R99P, RX1DAY, RX5DAY, SDII) compared to duration-based wet CEIs (e.g., R10mm, R20mm, R90D, R95D, R99D, WET, RX5DAYCOUNT, CWD). Throughout this paper, 'coupling strength' and 'correlation' are used interchangeably to represent the relationship between CEIs and fET. The coupling strength is generally positive and more pronounced for intensity-based wet CEIs compared to duration-based ones. Our findings reveal that the influence of watershed characteristics differs markedly across CEIs, highlighting the complex interactions between precipitation extremes, watershed properties, and ecosystem responses as measured by fET. Understanding these nuanced relationships is crucial for predicting and managing ecosystem responses to changing precipitation patterns.

### 3.1. Performance of GLEAM and ALEXI fET with CEIs

The metrics for GLEAM and ALEXI fET from 2001-2023 are shown in Fig. 1. GLEAM generally exhibits higher values, potentially overestimating ET in wet seasons over humid regions (57). Nevertheless, the correlation density plots for GLEAM and ALEXI sources exhibit similar patterns, generally agreeing on the direction of correlation (positive or negative) with CEIs with a few exceptions, Fig.2. Overall, GLEAM data shows stronger correlations, whereas ALEXI displays a wider range of correlation values. There are some notable differences between the two products, particularly in relation to correlation values for CEIs associated with precipitation amounts i.e., R90P and R99P (mean precipitation on days exceeding the 90th and 99th percentiles, respectively). The contrasting behaviors observed in R90P and R99P suggest that the two products may have different sensitivities to translation of extreme precipitation to ET, potentially due to variations in their underlying algorithms, input data, or spatial and temporal resolutions. These differences highlight the importance of considering dataset characteristics and temporal coverage when interpreting correlations between fET and CEIs. For the remainder of this paper, we will present results pertaining to GLEAM fET due to its long-term data availability. However, the analysis was also conducted using ALEXI fET and yielded consistent findings.

## 3.2. Casual and Predictive Relationship between CEIs-fET

A general pattern emerges across most CEIs: positive coupling between CEIs and fET for lags within ±2 months (Fig. 3), and negative coupling for lags exceeding 2 months. Cluster analysis identified mean lag times for causal and predictive relationships between CEIs and fET, grouping similar correlation-lag combinations, Table 2. Results predominantly reveal two clusters for most CEIs, with R20mm, R99P, and RX5Daycount exhibiting three clusters and higher mean time lags. These multi-timescale relationships reflect: (1) short-term atmospheric processes temporarily suppressing fET before rainfall or immediate fET increases with water availability; (2) medium-term ecosystem processes leading to soil moisture depletion after high fET periods; and (3) long-term effects on vegetation growth, water percolation, streamflow, and slope processes. The varying lag times and correlations reveal complex precipitation-ET dynamics, encompassing immediate and delayed effects, ecosystem memory, and feedback mechanisms.

In the short term (0-18 days), intensity-based CEIs such as R90P and R99P, which represent precipitation amounts exceeding the 90th and 99th percentiles, respectively, exhibit mean positive lags of ~ 0.168 and ~ 0.491 months, Fig.4. Both indices show negative correlations with fET, suggesting that extreme precipitation events temporarily suppress fET due to factors like waterlogged soils that limit plant water uptake. In contrast, less extreme precipitation events represented by RX1Day (the highest amount of precipitation received in a single day) and RX5Day (the highest amount over five consecutive days) demonstrate positive correlations with fET at lags of about ~0.139 months (~4 days) and ~0.164 months (~5 days), indicating that ecosystems can more readily utilize these moderate rainfall events, leading to increased fET after a short delay. However, RX1Day and RX5Day also exhibit negative lags of approximately -0.88 and -0.795 months, respectively, suggesting that decreases in fET tend to precede increases in wet periods by about a month. This pattern may reflect seasonal transitions where heightened humidity and cloud cover may temporarily suppress evapotranspiration before rainfall occurs. In terms of long-term effects, R99P shows positive correlations with lags of ~ -3.91 and ~ 5.3 months, indicating that watersheds may take this time to respond positively to water availability after intense precipitation, likely due to recovery from waterlogged conditions and gradual replenishment of soil moisture.

For duration-based CEIs, CWD (consecutive days with precipitation ≥1mm) shows a short positive lag of ~0.22 months (approximately 7 days), indicating a quick increase in fET following sustained rainfall as plants utilize newly available soil moisture, Fig.4. WET (days with precipitation ≥1mm) exhibit a longer positive lag of ~0.36 months (about 11 days), suggesting that the cumulative effects of multiple wet days take additional time to fully manifest in fET. This longer lag time for WET compared to CWD implies that the ecosystem response to precipitation differs between sustained and intermittent rainfall patterns. Both CWD and WD show a mean negative lag of ~-1.09 months (around -33 days), similar to RX1Day and RX5Day. Extreme precipitation events like R90P, R10mm, and R20mm generally display shorter mean lag periods alongside longer ones that may indicate seasonality. For instance, R10mm (days with precipitation ≥10mm) shows a positive lag of ~0.243 months (~7 days) with positive correlation and a negative lag of ~-1.89 months (around -57 days) with negative correlation. R20mm (days with precipitation ≥20mm) exhibits three clusters: a positive lag of ~0.618 months (about 19 days, positive correlation), ~5.56 months (~167 days, negative correlation), and a negative lag of ~-4.58 months (around -137 days, negative correlation). The short positive lags (7-19 days) indicate a rapid increase in fET following significant precipitation events, likely reflecting the immediate availability of water for vegetation uptake and subsequent transpiration. Conversely, the longer negative lags (-57 to -151 days) with negative correlations suggest that periods of high fET may precede drier periods. This could be due to multiple factors: vegetation depleting soil moisture reserves, seasonal patterns where periods of high fET (e.g., late spring/early summer) naturally precede drier periods (late summer/fall), or complex ecosystem responses to water availability. The presence of both positive and negative correlations at longer lag times, particularly for R20mm (5.56 months or ~ 167 days), indicates complex ecosystem memory effects that may relate to larger-scale climate patterns or oscillations, vegetation life cycles, or long-term soil moisture dynamics that influence both evapotranspiration and precipitation with different timing. Importantly, the distinct patterns for CWD and WET versus R10mm, R20mm and R90D highlight the importance of precipitation intensity, with more intense events showing more complex lag relationships, possibly due to their greater impact on soil water storage and runoff processes. These complex lag relationships underscore the importance of considering multiple timescales when analyzing ecosystem responses to extreme precipitation events.

### 3.3. Watershed Thresholds under Intensity-based CEIs – fET

Intensity-based CEIs exhibit a bimodal distribution in their correlation with fET, Fig.2. Positive coupling indicates that as intensity-based precipitation events increase, fET also increases, maintaining the soil-plant-atmosphere continuum. Conversely, negative coupling suggests that increased intensity-based precipitation events reduce fET, possibly due to waterlogging and potential run-off. The optimized RF model achieved RMSE=0.099, $R^2 = 0.959$ and accuracy = 91.35%. The SHAP analysis revealed, the coupling strength between these CEIs and fET is primarily influenced by SOC, GWT, and clay %, Fig.5. SOC is the primary driver, similar reasons stated above, with levels higher, i.e., ~ 131.44 ± 34.09 tonnes per hectare resulting in reduced coupling. The interaction between SOC and other variables was further analyzed, revealing that higher SOC levels are associated with lower GWT depths and lower clay %, Fig. This relationship can be explained by two factors. First, shallower water tables can create anaerobic conditions that slow down decomposition of organic matter, preserving more SOC in the soil. Second, sandy or loamy soils with lower clay content often have better aeration and drainage, which can promote root growth and organic matter input, Fig. GWT depth averaging 393.67 ± 172.30 cm strengthens the coupling. Deeper GWT allows for more water storage in the soil profile, enhancing the soil's capacity to buffer extreme precipitation events. The interaction of GWT with clay percentage was further analyzed Fig, revealing a clear demarcation: GWT > ~ 350 cm are associated with clay % above ~ 17 %. This can be attributed to the properties of clay-rich soils which tend to have lower hydraulic conductivity. This characteristic slows water movement through the soil profile. As a result, water takes longer to percolate and raise the water table, leading to deeper groundwater tables in these soils. Clay content above an average of ~ 23.57 ± 4.74% strengthens the coupling, while content below 11.04 ± 3.54% weakens it.

### 3.4. Watershed Thresholds under Duration-based CEIs – fET

For duration-based CEIs, the optimized RF model achieved RMSE=0.125, $R^2 = 0.80$, and accuracy = 80.50%. The SHAP analysis revealed that the watershed characteristics most significantly influencing this relationship are SOC, GWT, and permafrost %, Fig.6. Higher SOC enhances soil structure and water retention capacity (SWR), which increases plant-available water and infiltration. However, under extreme rainfall conditions, this can lead to waterlogging, thus reducing ET. Additionally, for the same SOC increase, soils with coarser textures present a larger increase in SWR than the finer soils, which may also present a decrease. This is because coarse soils, with naturally larger pores between particles, benefit more from added organic matter that helps develop additional small pores. Thus, the coupling strength between duration-based wet CEIs and fET decreases non-linearly with increasing SOC, Fig. Specifically, SOC levels below an average of ~19.01 (±11.09) tonnes per hectare hectare positively influence the coupling strength, enhancing fET. In contrast, SOC levels exceeding an average of ~ 70.84 (± 49.44) tonnes per hectare result in reduced coupling due to heightened waterlogging, which diminishes ET. Watersheds with negligible permafrost show no significant coupling strength. However, as permafrost percentage increases, with an average threshold of ~ 50.78% (±30.74%), the coupling strength becomes increasingly negative. This negative trend is attributed to these areas being in an energy-limited state, which inhibits ET. GWT depth averaging 394.21± 207.84 cm strengthens the coupling by providing greater storage through soil water percolation during heavy precipitation events and supporting evapotranspiration.

### 3.5. Coupling Strength of CEIs - fET with Terrestrial Biomes

The coupling strength varies significantly across 13 terrestrial biomes, Fig.7. Biomes 1 (Tropical and Subtropical Moist Broadleaf Forests), 2 (Tropical and Subtropical Dry Broadleaf Forests), 3 (Tropical & Subtropical Coniferous Forests), 6 (Boreal Forests), 7 (Tropical & Subtropical Grasslands, Savannas & Shrublands), 8 (Temperate Grasslands, Savannas & Shrublands) and 11 (Tundra) show either positive or negative coupling between CEIs and fET for different lag periods, suggesting more uniform responses to CEIs within these biomes. Specifically, Biomes 1, 2, and 3 exhibit positive coupling at smaller lag periods (-2 to 2) and negative coupling at higher lags (< -2 and > 2); Biomes 6 and 11 display negative coupling at smaller lags and positive coupling at higher lags. Exceptions include Biomes 1 and 11, which show transitions at lag = -2, and Biome 7, which transitions at lag = 4. In contrast, biomes 4 (Temperate Broadleaf & Mixed Forests), 5 (Temperate Conifer Forests), 10 (Montane Grasslands & Shrublands), 12 (Mediterranean Forests, Woodlands & Scrub), and 13 (Deserts & Xeric Shrublands) exhibit more complex behavior, with positive coupling at smaller lags and both positive and negative correlations at higher lags. This variability implies a transition in the coupling process, potentially reflecting shifts between water deficit and abundance conditions. Upon further analysis, transitions in coupling strength occur at higher lags for CEIs: R10mm, R20mm, R90P, R99P, RX1Day, RX5Daycount, and SDII. This suggests the impacts of large-scale climate patterns and seasonality on ecosystem water response, as discussed in section 3.1. Among all biomes, Biome 4 (Temperate Broadleaf & Mixed Forests) and Biome 8 (Temperate Grasslands, Savannas & Shrublands) show negative coupling for R90P and R99P at smaller lags, implying waterlogged conditions that temporarily reduce evapotranspiration which indicates their heightened sensitivity to wet extreme. The distinct response of these temperate biomes can be attributed to several factors. Firstly, the vegetation in these biomes has evolved to thrive in moderate moisture conditions, making it more susceptible to oversaturation. Secondly, temperate soils typically have higher organic content and clay fraction, enhancing water retention and prolonging saturation after heavy rainfall. Lastly, the generally shallower root systems of temperate vegetation, compared to tropical or arid counterparts, increase vulnerability to short-term soil saturation.

## 4. Conclusions

Climate extremes are projected to increase in frequency and intensity over the coming decades, making it essential to understand their relationship with evapotranspiration (ET), a key process linking water and carbon cycles. This study explores the complex interactions between Climate Extreme Indices (CEIs) and fractional evapotranspiration (fET) across various watersheds and terrestrial biomes. The research hypothesizes that these interactions are influenced by watershed characteristics and biome types. To investigate this, the study utilizes data from MERRA-2 climate indices, GLEAM, and ALEXI for evapotranspiration. While acknowledging limitations such as coarse resolutions and data uncertainties, the study leverages ALEXI's open-source nature, GLEAM's long-term global coverage, and MERRA-2's climatological foundation to provide valuable insights into wet climate extremes' impacts on ET. This comprehensive quantitative approach aims to identify thresholds for watershed characteristics that influence ecosystem productivity in response to climate extremes. The study also considers legacy effects, recognizing that dry or wet conditions in one year can significantly impact productivity in subsequent years. While the quantified thresholds for watershed characteristics may not be universally applicable and lack ground validation, this research serves as an initial effort to investigate and quantify thresholds essential for developing effective climate resilience strategies.

The analysis reveals multi-timescale relationships between CEIs and fET. Short-term positive coupling within ±2 months suggests rapid ecosystem responses to water availability, while negative coupling at longer lags indicates seasonality and broader ecosystem processes. Importantly, the study distinguishes between intensity-based and duration-based CEIs, as they exhibit distinct lag patterns and correlations with fET. For duration-based CEIs, soil organic carbon (SOC), permafrost %, and silt % significantly influence the CEI-fET coupling. Intensity-based CEIs are primarily affected by SOC, groundwater table depth, and clay %. The analysis reveals complex interactions between these factors and their thresholds, which determine the strength and direction of CEI-fET coupling. Across 13 terrestrial biomes, the relationship shows significant variations, with some biomes exhibiting uniform responses and others displaying more complex behaviors. The key highlights of the study are:

- Multi-timescale relationships between CEIs and fET reflect diverse ecosystem responses to climate extremes, with soil organic carbon (SOC) emerging as a primary driver for both duration-based and intensity-based CEI-fET coupling.
- Soil characteristics, including texture (silt and clay percentages) and groundwater table depth, significantly influence CEI-fET relationships, particularly in intensity-based CEIs.
- Among biomes, Temperate biomes, especially Broadleaf & Mixed Forests and Grasslands, exhibit heightened sensitivity to wet extremes.

These findings are crucial as they demonstrate the complex, time-dependent nature of the relationship between climate extremes and evapotranspiration. Future research should further explore similar studies considering higher-resolution data, ground-based validation, developing region-specific thresholds, and integrating additional ecological parameters. Exploring the impacts of land use changes and human interventions on potential cascading effects of climate extremes on interconnected ecosystems will also be crucial. Ultimately, this work lays the groundwork for more targeted studies and the development of adaptive management strategies to strengthen ecosystem resilience in the face of increasing climate extremes.

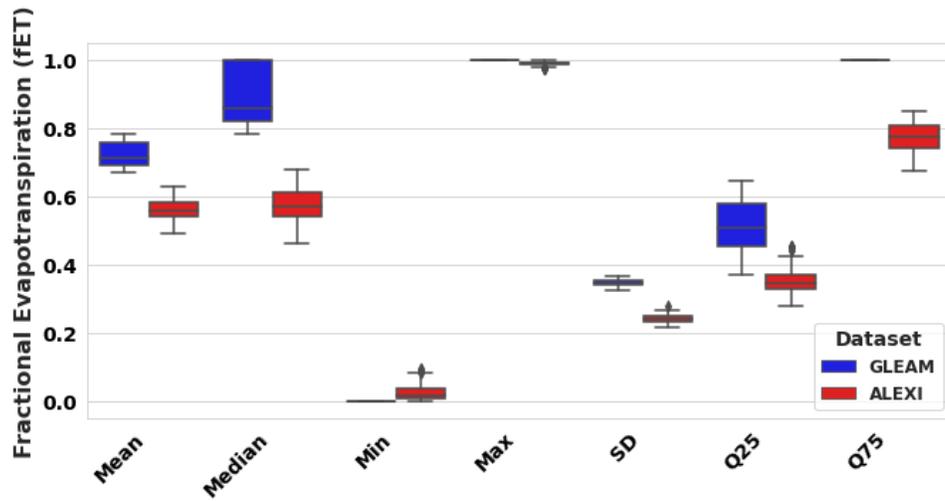

**Figure 1: Comparison of fractional evapotranspiration (fET) metrics for GLEAM and ALEXI datasets, 2001-2023.**

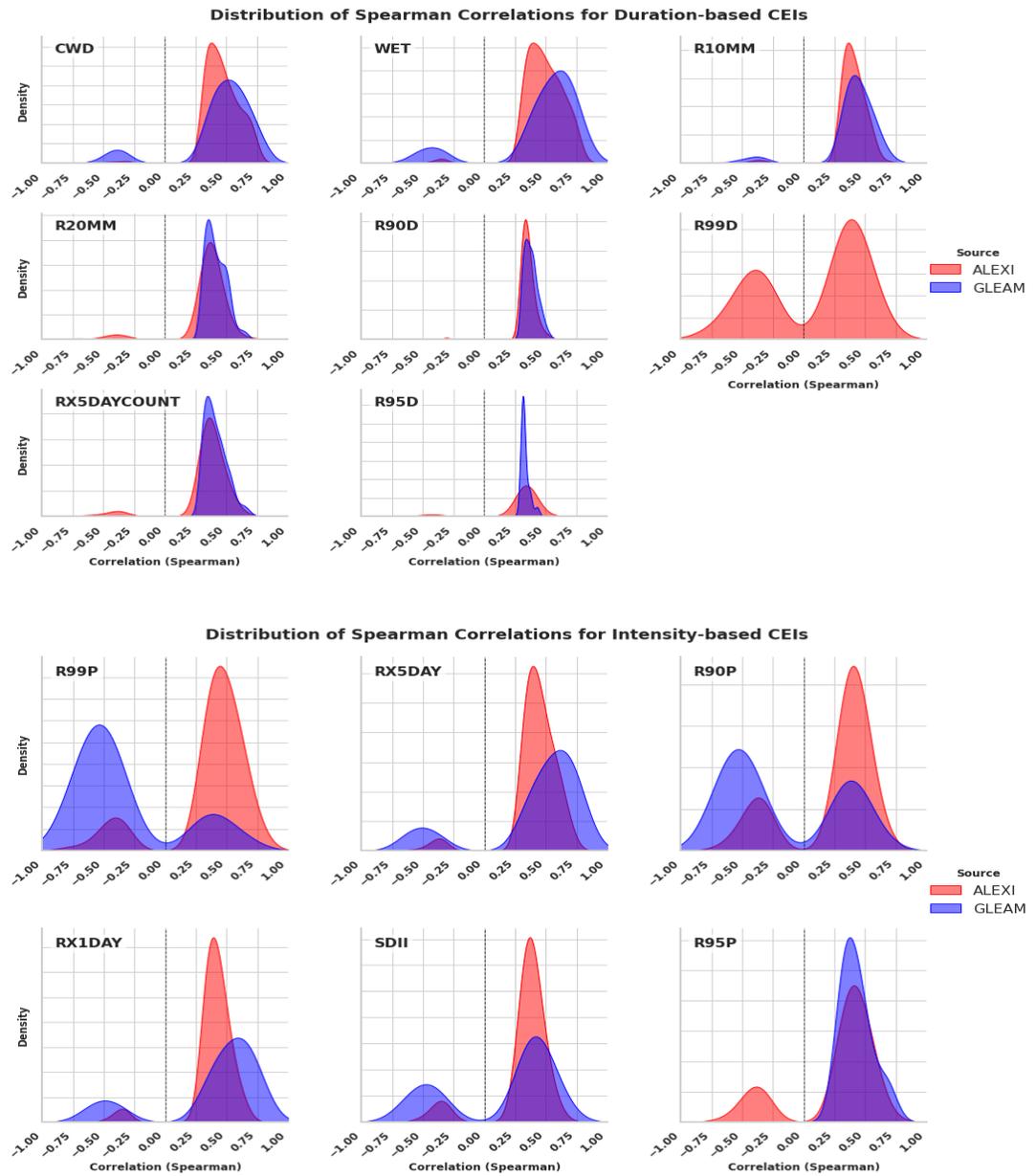

**Figure 2: Density plot showing the distribution of Spearman correlations between MERRA-2 Climate Extreme Indices (CEIs) and fractional evapotranspiration (fET) for ALEXI and GLEAM datasets. Red represents ALEXI data, and blue represent GLEAM data.**

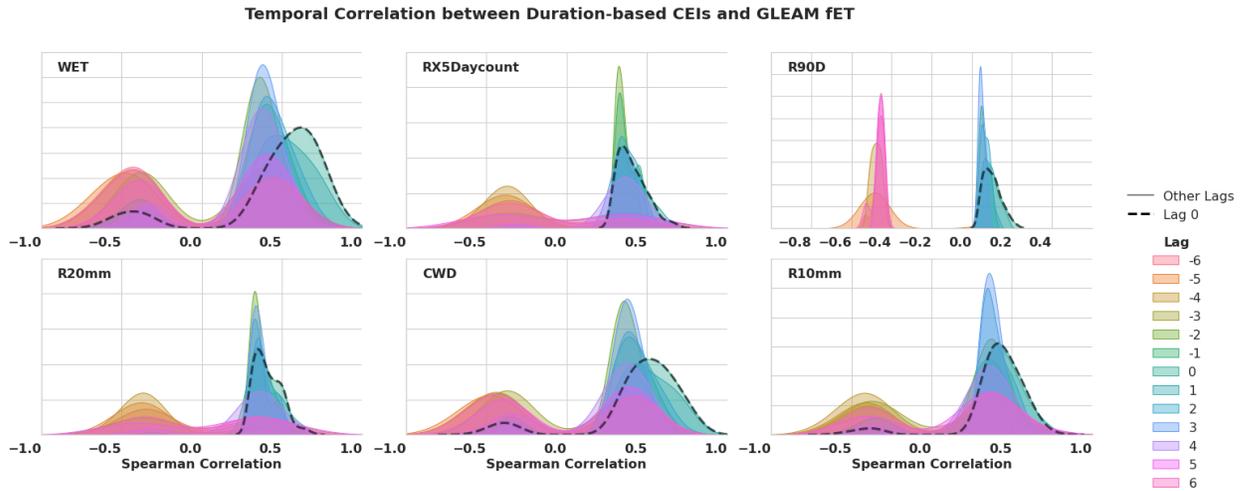

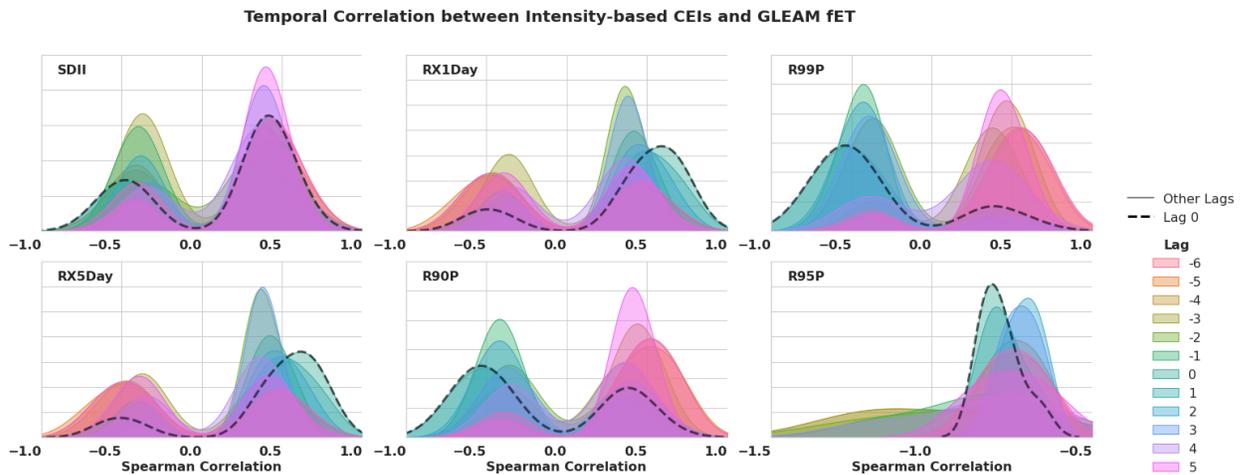

**Figure 3: Density plot showing the distribution of Spearman correlations between MERRA-2 Climate Extreme Indices (CEIs) and GLEAM fractional evapotranspiration (fET) at different time lags. The dotted black line represents lag=0, with different colors indicating various lag periods.**

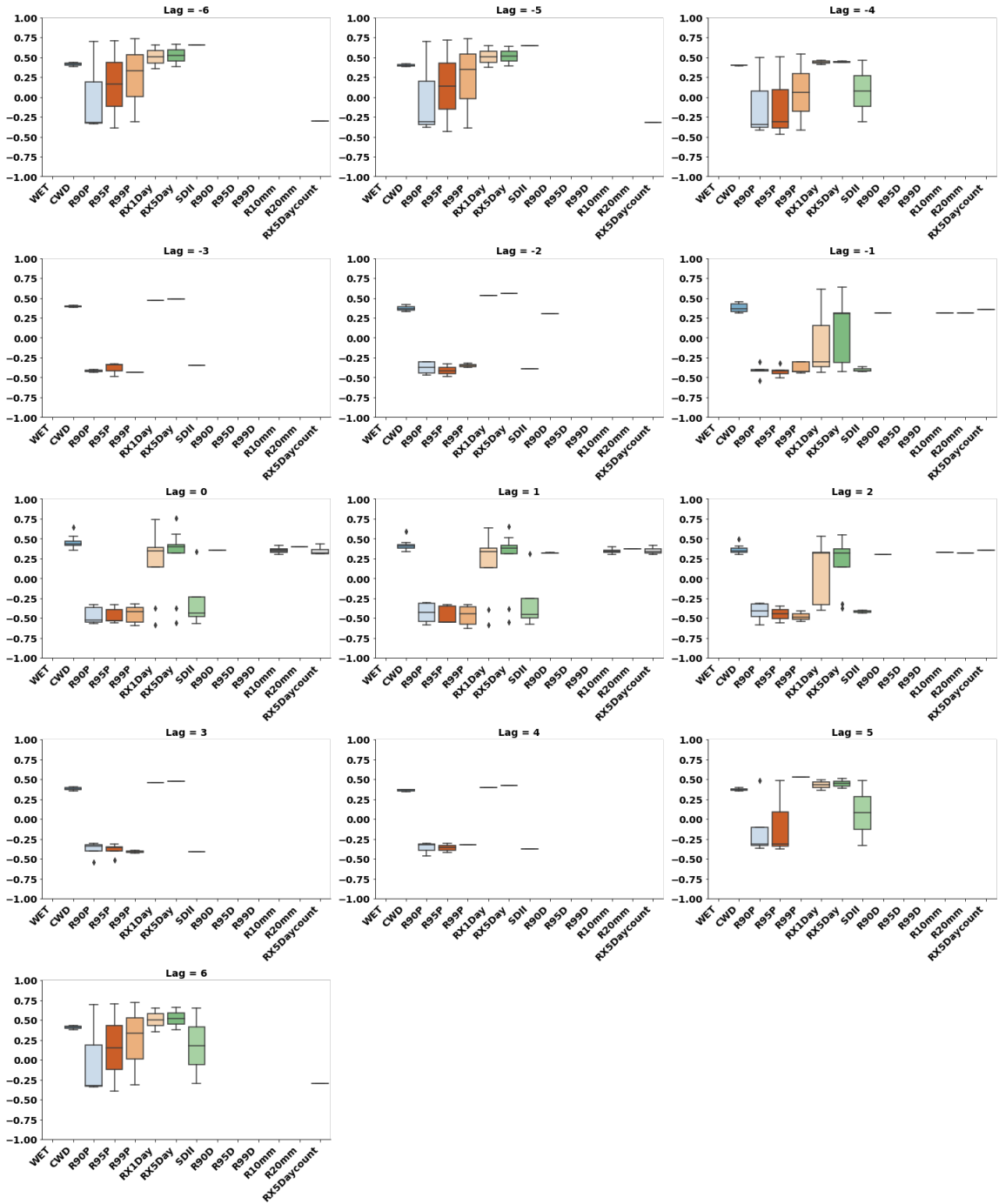

**Figure 4:** Box plots showing the distribution of correlation values for each CEIs with fET, displayed in subplots for different time lags.

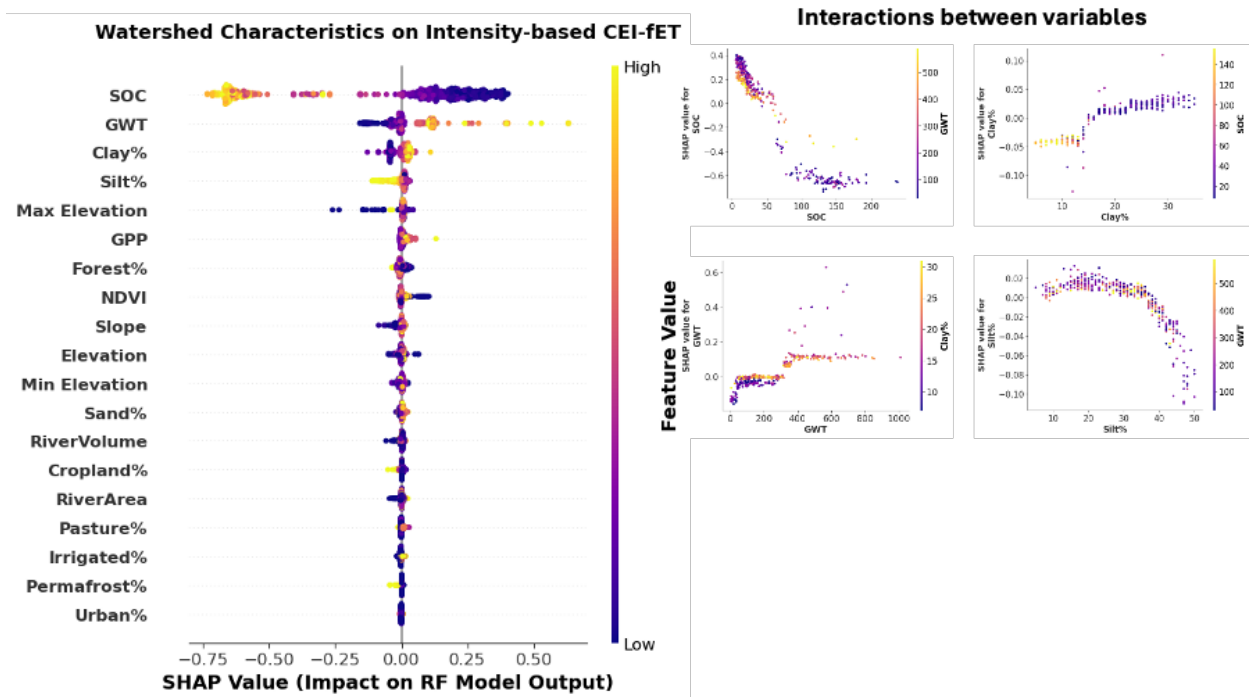

**Figure 5:** Feature importance analyzed using SHAP method to determine coupling strength between intensity-based CEIs and fET. Subplots illustrate interactions between significant variables.

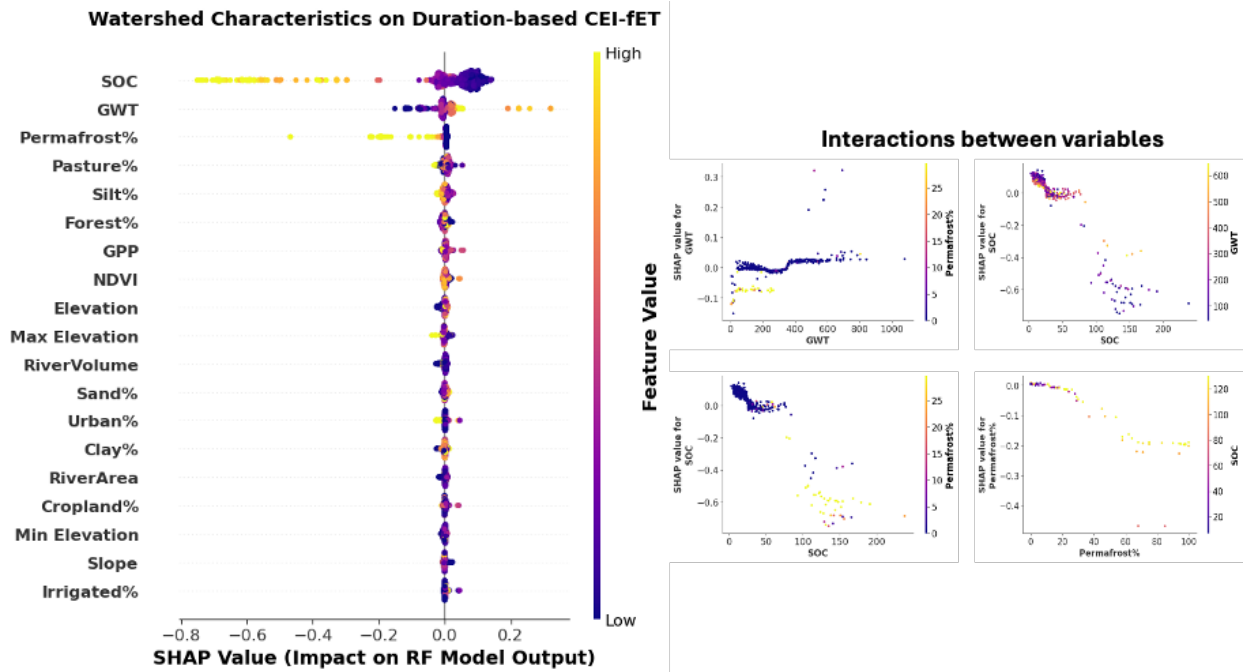

**Figure 6:** Feature importance analyzed using SHAP method to determine coupling strength between duration-based CEIs and fET. Subplots illustrate interactions between significant variables

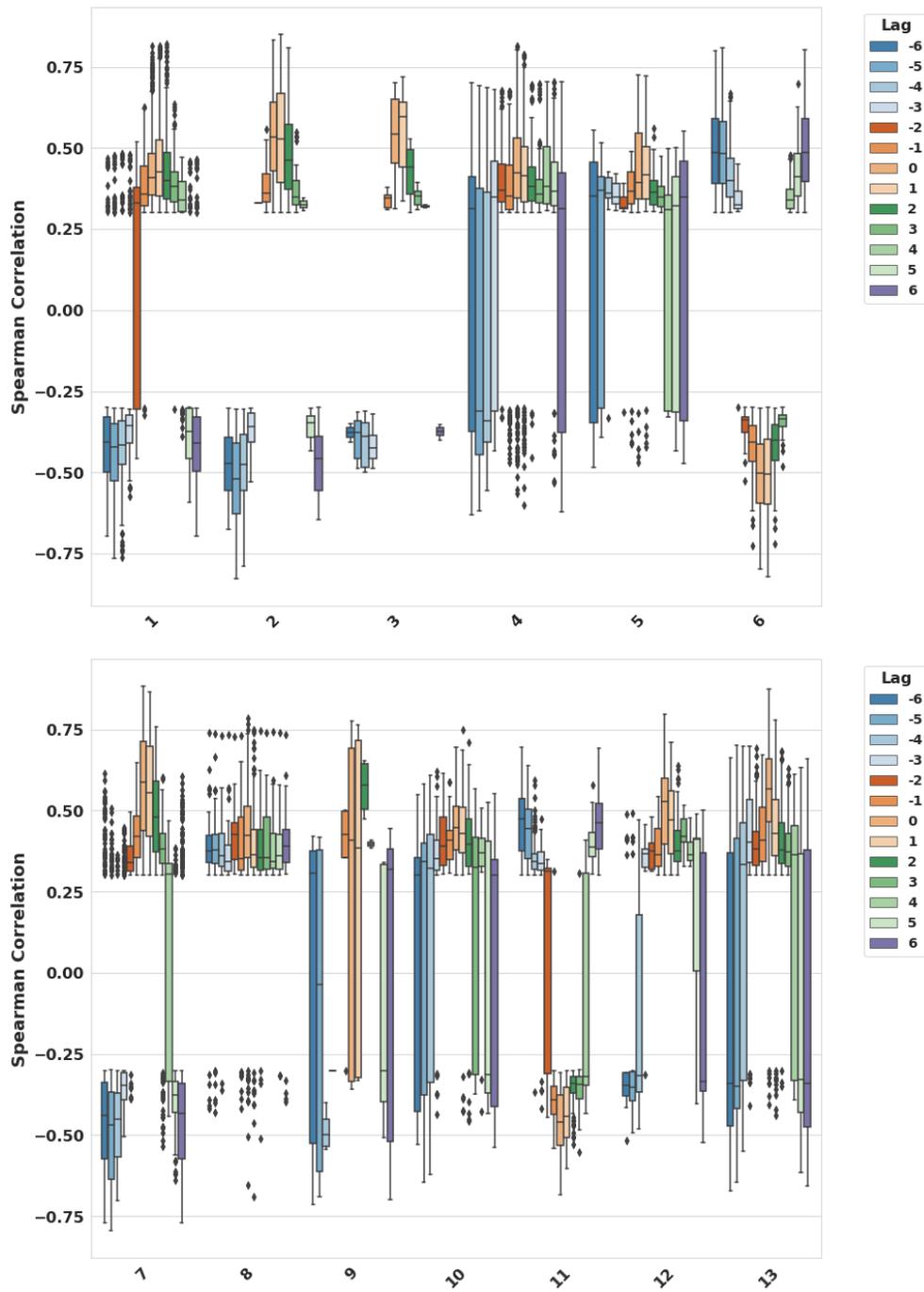

**Figure 7: Box plots showing the distribution of correlation between CEIs and fET at different lag times for various Biomes.**

**Data Availability:** The selected indices are included in the MERRA-2 extremes detection indices data product and are also available for visualization on the Global Modeling and Assimilation Office's Framework for Live User-Invoked Data (FLUID) webpage, https://fluid.nccs.nasa.gov/reanalysis/extreme_merra2/.

**Acknowledgments:**